\documentclass[prl,aps,twocolumn,notitlepage,floatfix,superscriptaddress,10pt]{revtex4-2}
\usepackage{physics}
\usepackage{amssymb}
\usepackage{amsmath}
\usepackage{mathtools}
\usepackage{braket}
\usepackage{notoccite}
\usepackage{dsfont}
\usepackage{tikz}
\usepackage{soul}
\usepackage{graphicx}
\usepackage{xcolor}
\usepackage[english]{babel}
\usepackage[colorlinks=true,urlcolor=blue,citecolor=blue,anchorcolor=red]{hyperref}

\newtheorem{theorem}{Theorem}
\newtheorem{corollary}[theorem]{Corollary}
\newtheorem{lemma}[theorem]{Lemma}

\DeclareMathOperator{\polylog}{polylog}

\usepackage[labelformat=simple]{subcaption}

\usepackage{ragged2e}
\DeclareCaptionJustification{justified}{\justifying}
\makeatletter
\def\maketitle{
\@author@finish
\title@column\titleblock@produce
\suppressfloats[t]}
\makeatother

\newcounter{SMsections}
\renewcommand{\theSMsections}{\Roman{SMsections}}
\DeclareRobustCommand{\SMsec}[1]{%
    \begin{center}
        \medskip
        \refstepcounter{SMsections}%
        \addcontentsline{toc}{section}{\theSMsections.\space#1}
        \textbf{\theSMsections.\quad\uppercase{#1}}
    \end{center}
}

\newtoggle{arXiv} 
\toggletrue{arXiv}      

\begin{document}
\newcommand{\orcidicon}[1]{\href{https://orcid.org/#1}{\includegraphics[height=\fontcharht\font`\B]{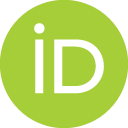}}}

\title{Near-Optimal Quantum Time Evolution Circuits \\via Provably Convergent Compression}

\author{Erenay~Karacan\orcidicon{0009-0000-1399-5084}}
\email{ekaracan@ethz.ch}
\affiliation{Department of Physics, ETH Zurich, Otto-Stern-Weg 1, 8093 Zurich, Switzerland}
\author{Isabel~Nha~Minh~Le\orcidicon{0000-0001-6707-044X}}
\affiliation{Technical University of Munich, CIT, Department of Computer Science, Boltzmannstraße 3, 85748 Garching, Germany}
\affiliation{Munich Center for Quantum Science and Technology (MCQST), Schellingstrasse 4, 80799 Munich, Germany}
\author{Matteo~D'Anna\orcidicon{0000-0002-9426-0377}}
\affiliation{Department of Physics, ETH Zurich, Otto-Stern-Weg 1, 8093 Zurich, Switzerland}
\affiliation{Institute for Theoretical Physics, ETH Zurich, Wolfgang-Pauli-Strasse 27, 8093 Zurich, Switzerland}
\author{Juan~Carasquilla\orcidicon{0000-0001-7263-3462}}
\affiliation{Department of Physics, ETH Zurich, Otto-Stern-Weg 1, 8093 Zurich, Switzerland}
\affiliation{Institute for Theoretical Physics, ETH Zurich, Wolfgang-Pauli-Strasse 27, 8093 Zurich, Switzerland}
\author{Christian~B.~Mendl\orcidicon{0000-0002-6386-0230}} 
\affiliation{Technical University of Munich, CIT, Department of Computer Science, Boltzmannstraße 3, 85748 Garching, Germany}
\affiliation{Munich Center for Quantum Science and Technology (MCQST), Schellingstrasse 4, 80799 Munich, Germany}
\affiliation{Technical University of Munich, Institute for Advanced Study, Lichtenbergstraße 2a, 85748 Garching, Germany}
\author{Ivan~Rojkov\orcidicon{0000-0001-7164-0265}}
\affiliation{Yale Quantum Institute, Yale University, New Haven, Connecticut 06520, USA}
\affiliation{Department of Physics, Yale University, New
Haven, Connecticut 06520, USA}

\begin{abstract}
Variational compression can significantly lower implementation overheads for encoding the time evolution of Hamiltonians into quantum circuits. However, they usually lack global convergence guarantees and well-established scaling behavior. In this work, we provide a recipe for choosing the initial point of such variational optimizations that guarantees convergence to a quantum circuit with near-optimal gate complexity $\mathcal{O}\left( N t \polylog(N t/\epsilon) \right)$ for all local and translationally invariant Hamiltonians. We demonstrate our method by encoding the \emph{globally controlled} time evolution of a Heisenberg antiferromagnet on a Kagome lattice. For $N = 48$ sites, evolution time $t=0.1$ and infidelity $\epsilon\approx1\%$, the controlled time-evolution circuit requires $960$ two-qubit B gates, for which we propose a straightforward implementation scheme for ion-trap setups. Thereby, our recipe extends digital quantum simulators toward system sizes and geometries that are challenging for classical computation.

\end{abstract}
\maketitle

Digital quantum simulation with quantum computers faces the challenge of transitioning from toy models to simulating large systems, not reachable by classical methods~\cite{Fauseweh2024}. Enabling such a leap requires algorithms that simultaneously fulfill three criteria~\cite{Georgescu_2014, Preskill2018NISQ}: (i) favorable asymptotic scaling of resource requirements with system size $N$ and target simulation error $\epsilon$, (ii) low implementation overhead beyond asymptotic scaling, such as minimal ancilla requirements and avoidance of costly oracle constructions \cite{Childs_2019, Low2019hamiltonian}, and (iii) rigorous performance guarantees, including quantified error bounds and, where applicable, guaranteed convergence.

Variational methods fulfill the low overhead criterion, but their scalability to large systems and reliability of their results are actively questioned~\cite{McClean_2018, Cerezo_2021}. By contrast, non-variational approaches, such as quantum phase estimation~\cite{Abrams_1999, Kitaev_1995, Cleve_1998} and adiabatic algorithms~\cite{Farhi_2000, Lidar_2007}, often come with rigorous scaling behavior and provable error bounds. While this makes non-variational methods appealing, they require approximating the continuous, unitary time evolution of a target Hamiltonian, using a discrete sequence of gates. This often represents a costly implementation bottleneck.

Many techniques have been developed for such a digital approximation of a Hamiltonian's time evolution. Among these, trade-offs are observed between the three aforementioned criteria: Trotter methods provide performance guarantees but exhibit non-optimal asymptotic scaling of resource requirements ~\cite{Su91, Lloyd1996Universal, Childs_2021}; quantum signal processing~\cite{Low_Chuang_2017} and qubitization~\cite{Low2019hamiltonian} achieve instead optimal \emph{query complexity} scaling $\mathcal{O}(N(t+\log (1/\epsilon)))$~\cite{Berry2015Optimal}, but have substantial implementation overheads from block encoding oracles~\cite{Childs_2019, Low2019hamiltonian, Karacan_2025}. Meanwhile, variational methods for approximating time-evolution unitaries ~\cite{Mitarai_2018, Pollmann_2021, Mizuta_2022, Dborin_2022, Conor_2023, Kotil_2024, Causer_2024, Le2025riemannianquantum, putterer2025high, Gibbs2025deepcircuit, danna2025circuitcompression2dquantum, mansuroglu2025preparationcircuitsmatrixproduct, ticc} exhibit low implementation overhead on quantum hardware, but typically lack established scaling behaviors and global convergence guarantees~\cite{Cerezo_2021, Bharti_2022}. Reconciling all three of the aforementioned criteria in a single algorithm has so far remained elusive.

\begin{figure*}
    \centering
    {\phantomsubcaption\label{fig:fig1_a}}
    {\phantomsubcaption\label{fig:fig1_b}}
    {\phantomsubcaption\label{fig:fig1_c}}
    {\phantomsubcaption\label{fig:fig1_d}}
    {\phantomsubcaption\label{fig:fig1_e}}
    \includegraphics[width=\linewidth]{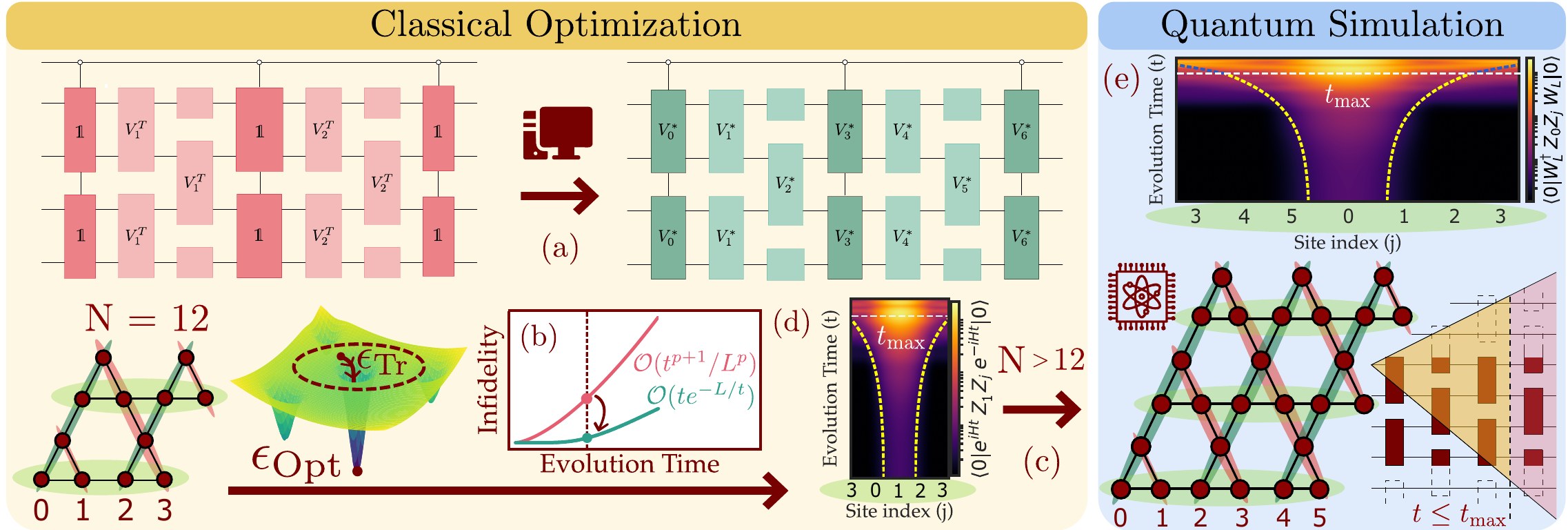}
    \caption{\textbf{Visualization of the proposed quantum simulation protocol.} (a) A local Ansatz circuit $W_L$ of depth $L$ is optimized to encode the controlled time evolution of time $t$ on a small quantum system (e.g., $N=12$ Kagome lattice) using a classical computer~\cite{ticc}. One uses an appropriate, $p^{\text{th}}$ order Trotterization of the time evolution with identities inserted as control layers~\cite{qetu, Babbush_2018, aqcf} for the initial point, which guarantees efficient convergence to a unitary with more favorable error scaling with respect to $L$ and $t$ compared to Trotter error, as shown in (b) in red for Trotterization and blue for optimized circuits. (c) Optimized gates can then be used to encode dynamics of the same Hamiltonian on a larger system, using a quantum computer. Approximation error on the larger system is well-bounded~\cite{Mizuta_2022}, if the targeted $t$ does not exceed $t_{\max}$, the time by which the Lieb-Robinson light-cone reaches the boundaries of the Ansatz system~\cite{LR}. The light cone spreading is visualized in (d) and (e) by evolving a product state (e.g., here $\ket{0}$) into an entangled state over time ($y$ axis in log-scale). This can be shown separately for each permutation class of the geometry (color-coded on the Kagome lattices). We plot equal-time correlations for one such permutation as a representative. }
    \label{fig:fig1}
\end{figure*}

In this Letter, we present a protocol for encoding the time evolution of translationally invariant (TI), local Hamiltonians into a quantum circuit with no ancilla or block-encoding overheads and near-optimal \emph{gate complexity} $\mathcal{O}( N t\polylog(N t/\epsilon) )$, which can also be extended to the controlled evolution~\cite{ticc}. Our method classically optimizes a local Ansatz circuit and transfers the optimized gates into larger, non-trivial systems, while keeping a quantified confidence interval on the expected evolution infidelity. A similar approach was proposed previously in Ref.~\cite{Mizuta_2022}. However, neither convergence guarantees for the variational optimization nor bounds on the resulting circuit depth were formalized. Here, we provide a recipe for choosing the initial point of the optimization and prove that it guarantees convergence to a unitary with the aforementioned favorable scaling. Thereby, we fulfill the low overhead, performance guarantee and favorable resource scaling criteria, all at once. A visual summary of our protocol is given in Fig.~\ref{fig:fig1}.

Convergence guarantees were investigated in quantum control before; however, they typically require the Ansatz to have an exponentially large number of parameters, covering the whole SU$\big(2^N\big)$~\cite{russell2016quantumcontrollandscapestrap, wiedmann2025convergencevariationalquantumeigensolver}. Here, we restrict our Ansatz to only a few layers, thus reducing the computational overhead of the classical optimization and avoiding the problem of barren plateaus in the optimization landscape~\cite{kiani2020learningunitariesgradientdescent, Cerezo_2021_2}.

\paragraph{Guaranteed convergence --} \citet{Haah_2021} proved that the system's time evolution governed by a local quantum Hamiltonian $H$ can be approximated with a local quantum circuit $G_L$, i.e. composed of two-qubit gates, of depth $L$, up to an approximation error
\begin{equation}
\label{eq:approx_err}
    \norm{G_L - e^{-i H t}} \leq \epsilon \quad \text{with} \quad \epsilon = \mathcal{O} \left(N \, t \, e^{- \Omega \left( (L/t)^{1/k} \right)}\right)
\end{equation}
for some $k \in \mathbb{N}^+$. The proof of this scaling uses, among others, the Lieb-Robinson bounds for information propagation in local quantum systems~\cite{LR}. In the particular case of a TI Hamiltonian $H$, the quantum circuit $G_L$ can also be taken TI~\cite{ticc}. In the following, we assume a finite system size $N$ and a finite Hilbert space dimension, hence $H$ is bounded.

The existence of such a unitary matrix $G_L$ motivates the use of a variational optimization scheme to find it with a local TI Ansatz $W_L(V)$ of depth $L$. Here, $V = (V^{(1)}, \dots, V^{(L)}) \in \text{SU}(4)^L$ is the parameter set of the Ansatz, where each $V^{(j)}$ is a fully parametrized two-qubit gate. As we take the two-qubit gate applied in each layer as fully parametrized, $G_L$ can be represented with a $W_L(V_{\text{opt}})$. 

The advantages of variational approaches have been previously demonstrated~\cite{Kotil_2024, Mizuta_2022, Pollmann_2021, Causer_2024, danna2025circuitcompression2dquantum}, where one can achieve approximation errors lower than that of Trotter circuits with comparable depth. In these protocols, one typically uses the trace norm as the cost function
\begin{equation}
\label{eq:cost_func}
    J_t(W_L(V)) \coloneqq -\text{Re}\big\{\text{Tr}[U_t^{\dagger} W_L(V)]\big\}
\end{equation}
of the optimization, where $U_t \coloneqq e^{-iHt}$ is the target. 

Such proposed variational methods typically do not provide a guarantee of convergence to an optimum $G_L$ satisfying Eq.~\eqref{eq:approx_err}, due to suboptimal minima (traps) and barren plateaus (flattening) in the optimization landscape. For efficient convergence, the workaround relies on heuristics to choose the optimization's initial point $V_0$, e.g. by starting from the Trotterized encoding of the target unitary \cite{Kotil_2024, danna2025circuitcompression2dquantum, ticc}.

Here, we instead propose a generic and provably successful strategy for achieving near-optimally scaling quantum circuits. We split the time evolution~$U_t$ into circuits of minimal depth $\Delta L$, approximating $U_{\Delta t}$, i.e. $t/\Delta t = L/\Delta L$. We prove that for \textit{sufficiently small} $\Delta t < t_{\text{crit}}$, our strategy for choosing the initial point $V_0$ guarantees the convergence of the optimization to a $G_{\Delta L}$, satisfying  Eq.~\eqref{eq:approx_err}. We then show that repeating these circuits to simulate time $t$, retains the near-optimal scaling of Eq.~\eqref{eq:approx_err}, despite the apparent sequential structure.

This minimal depth $\Delta L$ is the number of layers required to partition the Hamiltonian into local terms with disjoint supports. For nearest-neighbor spin systems this number is 2 in one dimension, 4 on the square lattice, and 6 on triangular or Kagome lattices.

\begin{theorem}
\label{theorem:theorem_1}
    Let $H$ be a local, TI Hamiltonian and $\Delta L$ its associated minimal circuit depth. There exists a time ${t_{\text{crit}}>0}$ such that $\forall\,\Delta t < t_{\text{crit}}$, a gradient-descent based optimization of 
    $J_{\Delta t}(W_{\Delta L}(V))$ is guaranteed to converge to $G_{\Delta L} = W_{\Delta L}(V_{\text{opt}})$, satisfying Eq.~\eqref{eq:approx_err}, if ${\frac{N}{2}\frac{\Delta L}{\Delta t} \norm{\log(V_0^{(j)})} \leq \norm{H}}$ for all $V_0^{(j)} \in V_0$.
\end{theorem}
In other words, choosing an arbitrary initial configuration~$V_0$, such that each ${V_0^{(j)} \in V_0}$ possesses a well defined principal matrix logarithm ${H_0^{(j)} \coloneqq \frac{i}{\Delta t}\log\!\big(V_0^{(j)}\big)}$ with $\frac{N \Delta L}{2} \norm*{H_0^{(j)}} \leq \norm{H}$, guarantees the optimizer's convergence to the best local approximation $G_{\Delta L}$ of depth $\Delta L$ with near-optimal scaling given in Eq.~\eqref{eq:approx_err} for a \emph{small enough} $\Delta t$. The proof of this theorem relies on the following lemma.

\begin{lemma}
\label{lemma:lemma_12}
 For the minimal circuit depth $\Delta L$, $\exists \, T>0$ such that for all local, TI Hamiltonians $H$ and $\forall \Delta t < T$ the cost function $J_{\Delta t}(W_{\Delta L}(V))$ in Eq.~\eqref{eq:cost_func} exhibits local convexity at the critical point $V_{\text{opt}}$. For all $\Delta t < T$, the convexity radius $R$ around $G_{\Delta L}$ is independent of $\Delta t$.
\end{lemma}

\begin{figure}
    \centering
    {\phantomsubcaption\label{fig:fig2_a}}
    {\phantomsubcaption\label{fig:fig2_b}}
    \includegraphics[width=1\linewidth]{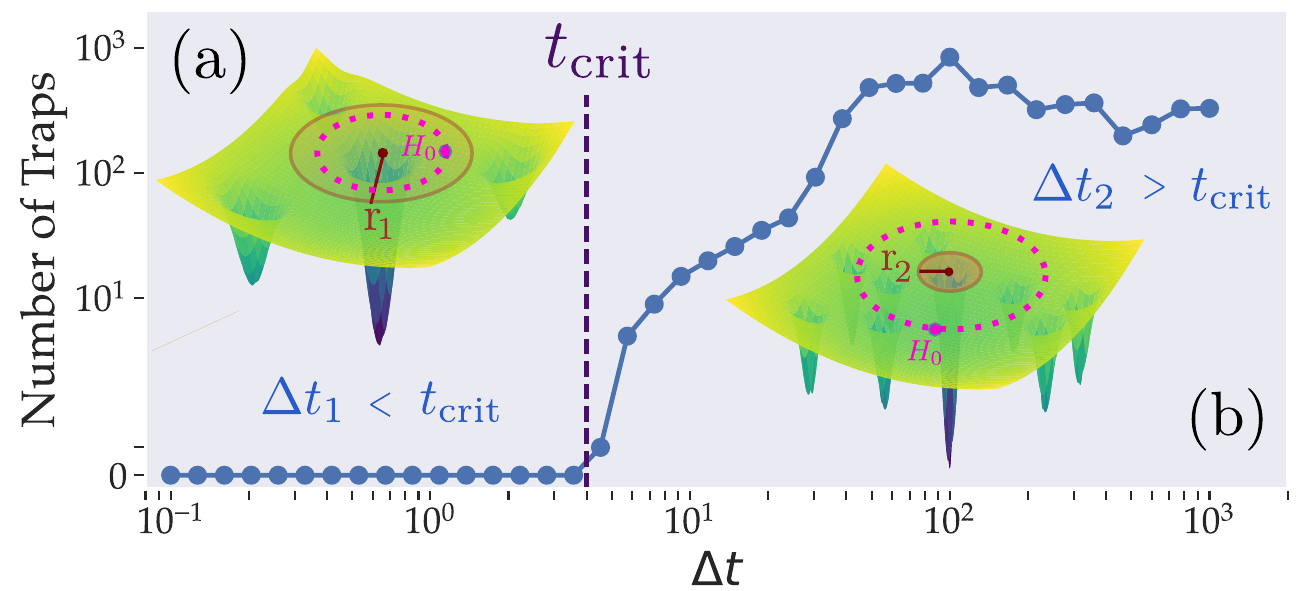}
    \caption{\textbf{Numerical demonstration of guaranteed convergence.} We fix $N = 6$ for the Ansatz, randomly choose a 2-local, TI Hamiltonian with $\norm{H} = 1$ on a fixed geometry with $\Delta L = 3$ permutation classes, set $U_{\Delta t} = e^{-i H \Delta t}$ as target and track convergence behavior for a randomized $H_0$, where $\norm{H_0} \leq 1$ and $e^{-iH_0 \Delta t}$ is the initial point of the optimization in unitary space. We repeat this for different $\Delta t$ values with 2000 random searches per $\Delta t$. For each $\Delta t$, a new randomized target $H$ is used. (a) The optimizer consistently converges to the same unitary in all random searches $\forall \Delta t < t_{\text{crit}} \sim 3$, because the set from which $H_0$ is randomized lies fully within the basin of attraction. (b) For $\Delta t > t_{\text{crit}}$, global convergence is not guaranteed anymore, and the optimizer occasionally gets stuck in local minima (traps), the number of which is displayed on the $y$-axis. }
    \label{fig:fig2}
\end{figure}

Local convexity implies that $\exists \, R > 0$ such that the cost function $J_{\Delta t}(W_{\Delta L}(V))$ is convex $\forall \, V$ satisfying ${\norm*{W_{\Delta L}(V)-G_{\Delta L}} \leq R}$ ~\cite{Bott, MilnorMorse, meyer2013introduction}. Importantly, $R$ is independent of the targeted evolution time~$\forall \Delta t < T$. We prove the Lemma in the Supplemental Material and proceed with the proof of Theorem~\ref{theorem:theorem_1}.

Let $H_0$ be a generator of the initial unitary ${e^{-i H_0 \Delta t}} = W_{\Delta L}(V_0)$. Thanks to local Lipschitz continuity of the exponential map and $H_0$ being Hermitian, it follows
\begin{equation} \label{eq:lipschitz_cont_exph0_GL}
    \norm{e^{-iH_0 \Delta t} -  G_{\Delta L}} \leq \Delta t \,\norm{H_0 - H_{\text{opt}}},
\end{equation}
where $H_{\text{opt}} \coloneqq \frac{i}{\Delta t} \log G_{\Delta L}$ in the principal matrix logarithm. Note that $H_{\text{opt}}$ is well defined via the principal matrix logarithm for sufficiently small $\Delta t$ (see Supplemental Material). Choosing $H_0$, with
\begin{equation}
\label{eq:conv_radius_ineq}
    \norm{H_0 - H_{\text{opt}}} \leq r \coloneqq \frac{R}{\Delta t}\,,
\end{equation}
implies that the initial unitary $e^{-iH_0 \Delta t}$ lies within the region of local convexity around $G_{\Delta L}$. Hence, we identify $r$ as a lower bound for the convexity radius in the Hamiltonian space with the property $r \propto\Delta t^{-1}$, i.e. this convexity radius only increases as $\Delta t$ decreases.

Taking any $H_0$ with $\norm{H_0} \leq \norm{H}$ will necessarily fulfill Eq.~\eqref{eq:conv_radius_ineq} for small enough $\Delta t < t_{\text{crit}}$, with 
\begin{equation}
\label{eq:t_crit}
    t_{\text{crit}} = \min \left\{T, \frac{\pi}{\norm{H}},  \frac{R}{2 \norm{H}} + \mathcal{O}\left( \frac{N \, e^{- \Omega ((\Delta L/\Delta t)^{1/k})}}{\norm{H}^2} \right) \right\}.
\end{equation}
The constraint $\norm{H_0} \leq \norm{H}$ can be satisfied by choosing any $V_0$, where each ${V_0^{(j)} = e^{-i H_0^{(j)} \Delta t} \in V_0}$ satisfies $\frac{N \Delta L}{2}\norm*{H_0^{(j)}} \leq \norm{H}$, as proposed in Theorem~\ref{theorem:theorem_1}. For details on this last step and the derivation of Eq.~\eqref{eq:t_crit}, please refer to Supplemental Material.

This way, warm starting the optimization from a product formula-based encoding of the time evolution, gains a rigorous guarantee with the condition on controlling the norm of $H_0$. Such Trotterization based initializations are direct ways of ensuring $\norm{H_0} \approx \norm{H}$ by construction, which is hereby proved to enable convergence for sufficiently short target times $\Delta t < t_{\text{crit}}$.

In practice, $t_{\text{crit}}$ can be determined numerically for the given Ansatz $W_{\Delta L}$, as shown in Fig.~\ref{fig:fig2}. Moreover, assuming linear accumulation of the error as we repeat the circuit of depth $\Delta L$~\cite{Childs_2021}, we conclude that the worst-case estimate of the approximation error is $\epsilon = \mathcal{O}\big(\frac{t}{\Delta t} \epsilon_0\big)$ for $\epsilon_0 = \mathcal{O} \left(N \, \Delta t \, e^{- \Omega \left( (\Delta L/\Delta t)^{1/k} \right)}\right)$. This leads to the following corollary.

\begin{corollary}
\label{corr:1.1}
    Repeating the circuit $W_{\Delta L}(V_{\text{opt}}) = G_{\Delta L}$ optimized for time step $\Delta t$, results in a circuit of depth $L$ prescribed by $\frac{L}{\Delta L} = \frac{t}{\Delta t}$. The depth $L$ scales with respect to accuracy $\epsilon$, system size $N$ and total evolution time $t$ as $L = \mathcal{O}(t \polylog(N \, t/\epsilon))$.
\end{corollary}

The approximation error $\epsilon$ of our construction fulfills the same favorable scaling as in Eq.~\eqref{eq:approx_err}. However, the obtained circuit depth $L$ may not scale optimally. We conjecture that our solution and the optimal, local approximation of $U_t$ differ by a polynomial degree $k$ in the poly-logarithmic factor $\polylog(N \, t/\epsilon)$.

\begin{figure}
    \centering
    \includegraphics[width=\linewidth]{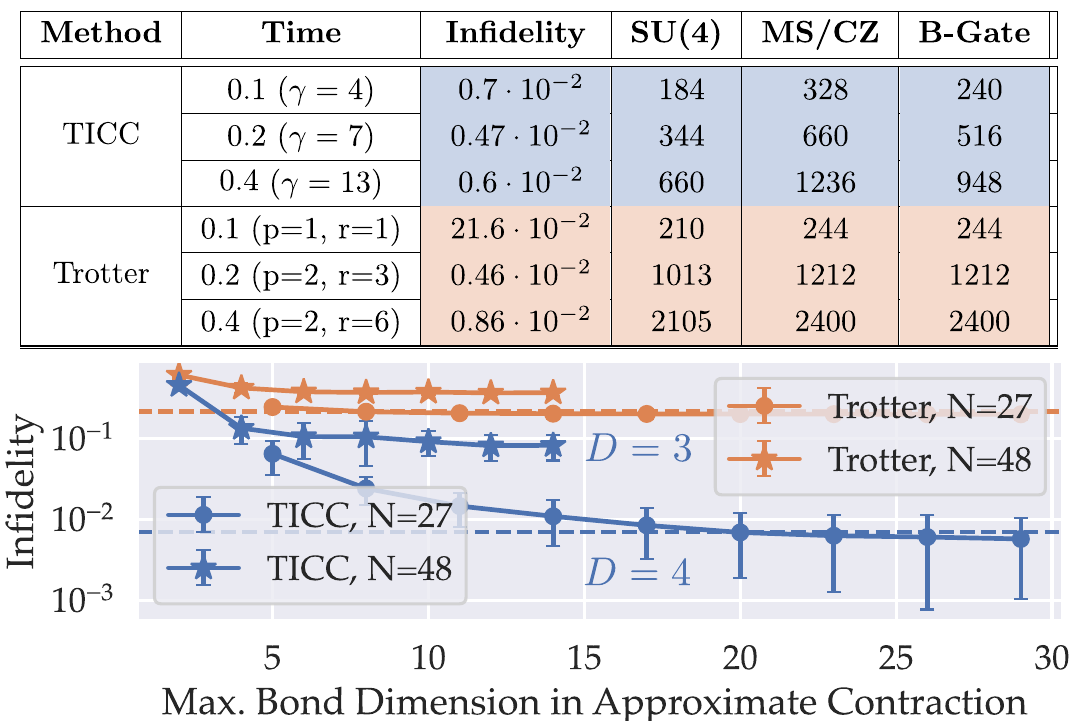}
    \caption{\textbf{Kagome lattice optimization results.} (Top) Evolution infidelity \eqref{eq:ev_infid} vs.\ gate counts needed to encode the \emph{controlled} time evolution of the HM in Eq.~\eqref{eq:HM}, on the $N = 12$ Kagome lattice. We benchmark TICC with control layers $\gamma$, against $p^{\text{th}}$-order Trotterization (of $r$ steps) with anti-commuting Pauli string insertion, as proposed by \cite{qetu, Babbush_2018}. Optimization with TICC yields arbitrary SU(4) gates, which we decompose into MS and B gates. (Bottom) PEPS verification of transferring optimized results to larger system sizes. We use a PEPS bond dimension $D = 4$ for $N = 27$, and $D = 3$ for $N = 48$. Dashed lines represent the evolution infidelities for the $N=12$ system, computed with exact diagonalization. PEPS contraction is performed approximately with the \texttt{hyper-compressed} method of \texttt{QUIMB}~\cite{quimb}. We average evolution infidelities of 20 random initial states with bond dimension 1, after evolving them for $t = 0.1$ (with TICC vs.\ Trotter) and computing their overlap with states evolved with finer Trotterization ($4^\text{th}$~order).}
    \label{fig:fig3}
\end{figure}

\paragraph{Optimization results on the Kagome lattice --} We demonstrate our method by optimizing \emph{globally controlled} time evolution circuits, a major bottleneck in many Hamiltonian simulation protocols \cite{nielsen_chuang, Georgescu_2014}.

We employ the recently proposed optimization framework called Translationally Invariant Compressed Control (TICC)~\cite{ticc}. It is based on extending the previous cost function in Eq.~\eqref{eq:cost_func} to the sum
\begin{equation}
    \Tilde{J}_t(W_L(V)) \coloneqq J_{-t}(W_L(V)) + J_{t}(W_{L-\gamma}(\Tilde{V})),
\end{equation}
where $\gamma>0$ is the number of \emph{control layers} and $\Tilde{V} \subset V$ are the gates of the reduced, \emph{uncontrolled} Ansatz. Controlling only the layers $V \backslash \Tilde{V}$ using an ancilla qubit implements a circuit where the ancilla state determines the evolution direction (i.e., forward $e^{-iHt}$ or backward $e^{iHt}$ in time), equivalent to the globally controlled evolution~\cite{Babbush_2018, qetu}. 

As a target, we use the Heisenberg Model (HM) in an arbitrary transverse field

\begin{equation}\label{eq:HM}
\begin{split}
    H = \sum_{\langle i, j \rangle} \sigma^X_i \sigma^X_j + \sigma^Y_i \sigma^Y_j + \sigma^Z_i \sigma^Z_j + \sum_{i} 3\sigma^X_i - \sigma^Y_i + \sigma^Z_i
\end{split}
\end{equation}
on the Kagome lattice with periodic boundaries, where $\langle i, j \rangle$ runs over the nearest neighbors and $\sigma^X_i, \sigma^Y_i, \sigma^Z_i$ are the Pauli operators acting on site $i$. This is a paradigmatic, strongly frustrated system whose ground-state properties in the thermodynamic limit remain unresolved~\cite{Kagome_1, Kagome_2, Kagome_3}.

To alleviate the computational cost, we modify the cost function in Eq.~\eqref{eq:cost_func} using random state-vector samples instead of full unitaries~\cite{caro2023out,zhang2024scalable}
\begin{equation}\label{eq:C_hst}
    C_t(W_L(V)) \coloneqq  1 - \mathbb{E} \left[ \abs*{\braket{ v | U_t^{\dagger} W_L(V) | v}}^2 \right]_{v\sim\text{Haar}},
\end{equation}
where $v$ is sampled from the uniform distribution, induced by the global Haar measure. This is equivalent to the earlier introduced trace norm cost in Eq.~\eqref{eq:cost_func} for large enough sampling~\cite{Nielsen_2002, Khatri_2019}.

Motivated by the aforementioned insights about convergence, we choose an appropriate Trotterization for initializing the uncontrolled layers and insert identities $\mathds{1}$ to initialize the control layers. By construction, this ensures $\norm{H_0} \approx \norm{H}$ for both circuits $W_L$ and $W_{L-\gamma}$, hence guaranteeing convergence for small $\Delta t$. Such an initialization is particularly useful for TICC, as one does not have access to a straightforward Trotterization of the full circuit for generic $\gamma$.

Moreover, we use bootstrapping of shallow layers $\Delta L$, optimized for small time steps $\Delta t$, to encode larger times. In this work, we report that this bootstrapping strategy yields efficient convergence for arbitrary circuit depths $L$ and leave the extension of Theorem \ref{theorem:theorem_1} to arbitrary $L$ for future work.

To quantify the performance of each method, we evaluate the evolution infidelity
\begin{equation}
\label{eq:ev_infid}
    1 - \mathbb{E} \left[ \abs*{\braket{v | U^{\dagger}_t U | v} }^2 \right]_{v\sim\text{Haar}}
\end{equation}
over randomized state-vectors $v$, where $U$ is the approximate unitary (TICC and Trotter circuits).

Our classical optimization results for system size $N = 12$ are presented in Fig.~\ref{fig:fig3}. As proposed previously \cite{Mizuta_2022, Kotil_2024, ticc}, these optimized gates can be transferred to larger systems $\Tilde{N} > N$ as long as the evolution time is below $t_{\text{max}} = \mathcal{O}(N^{2/\Delta L}/v_{\text{LR}})$ where $v_{\text{LR}}$ is the Lieb-Robinson velocity of the Hamiltonian \cite{LR}. Importantly, the increase in the approximation error is predictable and, in the worst case, linear in $\Tilde{N}$~\cite{Mizuta_2022}. For targeted $t<t_{\text{max}}$, we can use our gates, optimized on the $N=12$ system, to encode evolution dynamics of e.g. the $\Tilde{N}=48$ system on a quantum computer.

To investigate the transferability of our optimization results, we run projected entangled-pair states (PEPS)~\cite{verstraete2004renormalizationalgorithmsquantummanybody} simulations on sizes $\Tilde{N}=27$ and $\Tilde{N}=48$ (Fig.~\ref{fig:fig3}, bottom). The error we recover on the $\Tilde{N}=27$ system for $t = 0.1$ is similar to that from $N = 12$. Hence, we conclude that our optimization results for $t = 0.1$ fulfill $t < t_{\max}$ and can be transferred to even larger systems with the increase in evolution error staying bounded up to Lieb-Robinson error~\cite{LR, Mizuta_2022}. We attribute the higher infidelities we read from $\Tilde{N}=48$ system in Fig.~\ref{fig:fig3} to limitations of PEPS and computational resources. Additionally, we verify this transferability using Pauli Propagation framework \cite{rudolph2025paulipropagationcomputationalframework}, the results of which agree with PEPS simulations (see Supplemental Material).

\paragraph{TI Ansatz composed of B Gates --} Our optimizations yield arbitrary SU(4) gates as optimized results. In practice, such arbitrary gates are implemented using single-qubit rotations and three controlled-Z (CZ) or Mølmer–Sørensen (MS) gates, which are the native entangling operations for superconducting qubits~\cite{dicarlo_demonstration_2009} and trapped ions~\cite{ASENS_1927_3_44__345_0, M_lmer_1999, RevModPhys.75.281}, respectively. The decomposition of an arbitrary SU(4) gate can however be reduced to two fixed-parameter two-body operations known as the B gate~\cite{Zhang_2004}
\begin{equation}
\label{eq:B_gate}
    U_{\text{B}} = \exp\left(-i\frac{\pi}{4}\Big(\sigma_1^X \sigma_2^X + \frac{1}{2} \sigma_1^Y \sigma_2^Y\Big)\right),
\end{equation}
thus potentially reducing the total number of gates up to 1/3 (see Fig.~\ref{fig:fig3}). This operation has been recently demonstrated in a superconducting qubit architecture~\cite{Chen2025EfficientTwoQubit}. Here, we propose a hardware-native approach to implementing a B gate in ion-trap platforms.

Driving two target ions with bichromatic laser fields, detuned by $\delta$ from resonance, implements the state-dependent force (SDF) Hamiltonian (see End Matter). The system's evolution for a time $t=2 \pi/\delta$ under this Hamiltonian results in the unitary 
\begin{equation}
\label{eq:U_SDF}
    U_{\text{SDF}}(\delta, \Phi) =  \exp \left( -i \pi \, \frac{\Omega^2 \eta^2}{\delta^2} \, \sigma_1^{\Phi}\sigma_2^{\Phi} \right).
\end{equation} 
Here, ${\Phi \coloneqq (\varphi_b + \varphi_r)/2}$ is the average phase of the blue and red detuned lasers, $\Omega$ is the Rabi frequency, $\eta$ is the Lamb-Dicke parameter and ${\sigma_i^{\Phi} \coloneqq \cos(\Phi) \sigma^X_i + \sin(\Phi) \sigma^Y_i}$ is the modified spin operator acting on qubit $i$ ~\cite{ASENS_1927_3_44__345_0, M_lmer_1999, RevModPhys.75.281}.

The B gate in Eq.~\eqref{eq:B_gate} can then be implemented using 
\begin{equation}
\label{eq:B_SDF}
\begin{split}
    U_{\text{SDF}}(\delta,\Phi) \, 
    U_{\text{SDF}}(\delta,-\Phi) \,\,\,\text{with}\, 
    \left\{\begin{matrix*}[l]
        \Phi =  \frac{1}{2} \arccos(\sqrt{2}-1)\,, \\[8pt]
        \delta=\frac{4}{\sqrt{3}}\,\Omega\,\eta\,,
    \end{matrix*}\right.
\end{split}
\end{equation}
which is equivalent to a multi-loop MS gate construction, previously used to mitigate low-frequency noise channels of an MS gate. By setting the average phase $\Phi$ as given in Eq.~\eqref{eq:B_SDF}, this construction leads to a B gate, instead. See End Matter for the derivation.

Multi-loop MS gates have been previously demonstrated to an average infidelity as low as ${10^{-3}}$~\cite{Hayes_2012, Choi_2014, Shapira_2023}. We therefore expect no significant reduction in the B gate's fidelity, which we verify through numerical simulations (see End Matter).

\paragraph{Discussion and Conclusion --} 
In this work, we provided a provably successful strategy for variational optimization of quantum circuits, encoding time evolutions of local, TI Hamiltonians with nearly optimal scaling depth complexity.

Motivated and guided by these insights about the convergence behavior, we optimized the globally controlled time evolution of the Kagome lattice Heisenberg antiferromagnet. We verified the transferability of our optimized circuits to larger system sizes through PEPS simulations and we report that our method enables encoding the controlled time evolution of Kagome antiferromagnet with $48$ sites at $t=0.1$ with evolution infidelity $\sim1\%$, using 960 B gates in total (476 in the longest path, see Supplemental Material for details on gate synthesis). This low gate count on such a non-trivial system can help digital quantum simulation enter the domain that is challenging for classical computation, as the primary bottleneck for some of the most promising quantum algorithms is the high implementation cost of the controlled evolution \cite{CiracCosineFilter, rpe, qetu, aqcf}.

Our strategy for encoding the time evolution for an arbitrary time $t > 0$, although scaling near-optimally as given in Eq.~\eqref{eq:approx_err}, presumably does not find the best local approximation for layers $L$ as we naively repeat the layers optimized for $\Delta t \ll t$ sequentially. This prediction is confirmed by our numerical investigation of the Kagome lattice, as one can lower approximation errors further by running an optimization for all of the $L$ layers, with such linearly repeated circuits as the starting point (i.e., through bootstrapping). Our empirical observation is that such a bootstrapping strategy may guarantee convergence to the best local approximation for any depth $L$ and time $t > 0$. Investigating such convergence behavior is left for future work.

\paragraph{Acknowledgments --} EK thanks Ignacio Cirac and Jonathan Home for insightful discussions regarding the conceptual development of the B gate proposal. IL and CM thank the Munich Center for Quantum Science and Technology for support. IR acknowledges funding from the Swiss National Science Foundation (Postdoc.Mobility Grant No.~P500-2$\_$235497).

\bibliography{bibliography.bib}
\clearpage
\onecolumngrid
\section{End Matter}
\twocolumngrid

\begin{figure}[b]
    \centering
    \includegraphics[width=0.99\linewidth]{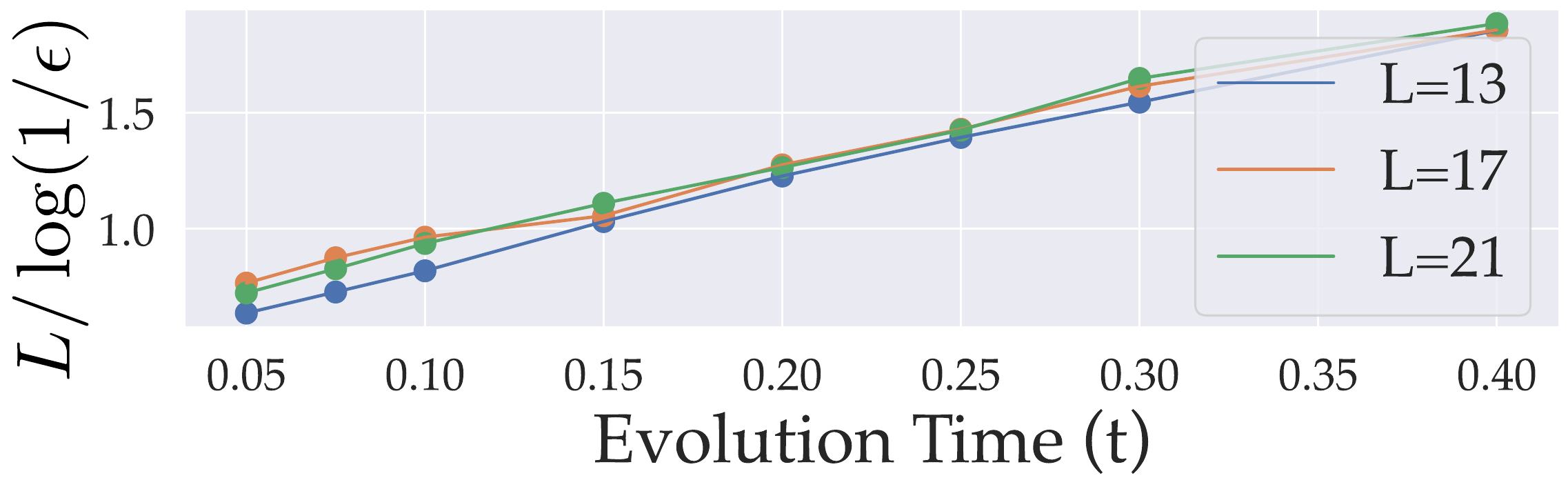}
    \caption{Scaling of the controlled time evolution operator (optimized via TICC), with respect to the evolution time, for various numbers of Ansatz layers $L$, where the number of control layers is set to a constant $\gamma = 4$ for all $L$ values. The inverse accuracy behavior $1/\log(1/\epsilon)$ is plotted and rescaled with $L$ to show the data points collapsing.}
    \label{fig:fig4}
\end{figure}

\paragraph{Scaling of the controlled evolution --} Although controlled time evolution Ansatz falls outside of the scope of Theorem~\ref{theorem:theorem_1} and Corollary~\ref{corr:1.1}, we numerically show that it can satisfy the same scaling behavior when the number of control layers $\gamma$ is kept constant for increasing $t$. This is demonstrated in Fig.~\ref{fig:fig4}, for the HM \eqref{eq:HM} on a one-dimensional geometry.

Control layers split the optimal unitary into ${G_L = G^1_L G^2_L \dots}$ such that for each sub-unitary $G^i_L$ there exists an anti-commuting control pair $K_{i-1} G^i_L K_{i} = -G^i_L$. The number of such unitaries is independent of $t$, because it only depends on the Hamiltonian structure $H$, as explained in Ref. \cite{ticc}. This fact explains the scaling behavior reported in Fig. \ref{fig:fig4} for constant $\gamma$. Despite that, we use bootstrapping of circuits optimized for short times, which results in an inevitable increase of $\gamma$, as $t$ is increased (see Fig. \ref{fig:fig3}). Providing an initial point for the control layers, which can enable efficient convergence of TICC for deep Ansätze at large evolution times, remains to be a difficult task.

\paragraph{B gate derivation --} Our B gate implementation relies on state dependent forces. Both $U_B$ and $U_{\mathrm{SDF}}$ commute with the parity operator 
$\sigma^Z_1\sigma^Z_2$, as their generators are linear 
combinations of $\sigma_1^X \sigma_2^X$, $\sigma_1^Y \sigma_2^Y$, and $\sigma_1^X \sigma_2^Y+\sigma_1^Y \sigma_2^X$ (see Eq. \eqref{eq:B_gate} and Eq. \eqref{eq:U_SDF}), all of which preserve parity. Therefore, they are block-diagonal in the even $\{\ket{00},\ket{11}\}$ and odd $\{\ket{01},\ket{10}\}$ subspaces with each block represented as a $2 \times 2$ unitary.

We start by representing the B-gate propagator $U_B$ in Eq.~\eqref{eq:B_gate} in the even and odd subspaces. The two-body Pauli operator $\sigma_1^X \sigma_2^X$ maps $\ket{00}$ to $\ket{11}$, while $\sigma_1^Y \sigma_2^Y$ realizes $\ket{00}\mapsto-\ket{11}$. We can therefore map these operators $\sigma_1^X \sigma_2^X \mapsto  X_\mathrm{even}, \, 
\sigma_1^Y \sigma_2^Y \mapsto -X_\mathrm{even}$, where $X_\mathrm{even}$ is the Pauli-$X$ operator in the even subspace $\{\ket{00},\ket{11}\}$. Similarly, in the odd subspace $\{\ket{01},\ket{10}\}$, we have $\sigma_1^X \sigma_2^X\,\ket{01} = \ket{10}$ and $\sigma_1^Y \sigma_2^Y\,\ket{01} = \ket{10}$ so that $\sigma_1^X \sigma_2^X \mapsto X_\mathrm{odd}, \, \sigma_1^Y \sigma_2^Y \mapsto X_\mathrm{odd}$, where now $X_\mathrm{odd}$ is the Pauli-$X$ operator in the basis $\{\ket{01},\ket{10}\}$ (the ``odd'' and ``even'' subscripts are superfluous but are kept for clarity). Hence, we get
\begin{equation} \label{eq:U_B_even_odd}
\begin{split}
    U_B = \exp\!\left(-i\frac{\pi}{8} X_\mathrm{even}\right)\oplus
    \exp\left(-i\frac{3\pi}{8}X_\mathrm{odd}\right).
\end{split}
\end{equation}

\noindent
Let us now consider the echoed SDF sequence
\begin{equation}
    B(\delta,\Phi) \equiv U_{\text{SDF}}(\delta,\Phi)\,U_{\text{SDF}}(\delta,-\Phi)
    = e^{-i\theta H_+} e^{-i\theta H_-},
\end{equation}
with $\theta = \pi \Omega^2 \eta^2 / \delta^2$ and 
\begin{equation}
\begin{split}
    H_\pm = \,&\cos^2\Phi\,\sigma_1^X \sigma_2^X + \sin^2\Phi\,\sigma_1^Y \sigma_2^Y \\ 
    &\pm \cos\Phi\sin\Phi\,(\sigma_1^X \sigma_2^Y+\sigma_1^Y \sigma_2^X).
\end{split}
\end{equation}

Given the parity symmetry mentioned earlier, we can write $B=B_\text{even} \oplus B_\text{odd}$ with $B_\text{even}$ and $B_\text{odd}$ being again $2 \times 2$ unitaries parametrized by $\Phi$ and $\theta$.
In the odd subspace $\{\ket{01},\ket{10}\}$, the Pauli cross-terms act trivially, $(\sigma_1^X \sigma_2^Y+\sigma_1^Y \sigma_2^X)\ket{01}=(\sigma_1^X \sigma_2^Y+\sigma_1^Y \sigma_2^X)\ket{10}=0$, so that
\begin{equation}
\label{eq:B_odd}
    B_\text{odd}(\theta) = \exp\!\left(-i\,2\theta\, X_\mathrm{odd}\right).
\end{equation}
The independence of $B_\text{odd}$ from $\Phi$ allows us to fix ${\theta=3\pi/16}$ (cf. Eq.~\eqref{eq:U_B_even_odd}).

In the even subspace, the cross terms act as following: $(\sigma_1^X \sigma_2^Y+\sigma_1^Y \sigma_2^X)\,\ket{00} = 2i\ket{11}$, leading to
\begin{equation}
\begin{split}
    B_{\mathrm{even}}(\theta,\Phi) = 
    &\exp\!\left(-i\theta\,\big[\cos(2\Phi)\,X_\mathrm{even}+\sin(2\Phi)\,Y_\mathrm{even}\big]\right) \\
    &\exp\!\left(-i\theta\,\big[\cos(2\Phi)\,X_\mathrm{even}-\sin(2\Phi)\,Y_\mathrm{even}\big]\right).
\end{split}
\end{equation}
where $Y_\mathrm{even}$ is the Pauli-$Y$ operator defined in the subspace $\{\ket{00}, \ket{11}\}$. Simplifying further, we get
\begin{equation}
    B_{\mathrm{even}}(\theta,\Phi) = 
    \begin{pmatrix}
        d & -i\,t \\
        -i\,t & d^*
    \end{pmatrix}
\end{equation}
where $d=\cos^2\theta-\sin^2\theta\,e^{-i4\Phi}$ and $t=\sin(2\theta)\cos(2\Phi)$. Under global Pauli-$Z$ rotations $R_z(\lambda)\otimes R_z(\lambda)$ with ${\lambda = \tfrac{1}{2}\arg d}$, the propagator in the odd sector remains unchanged while the even one gets transformed to 
\begin{equation}
\label{eq:B_even}
    |d|\,\mathds{1}_\mathrm{even} - i\,t\,X_\mathrm{even}
    = \exp\!\left(-i\,\beta \, X_\mathrm{even}\right),
\end{equation}
with $\sin(\beta) = \sin(2\theta)\cos(2\Phi)$. Such global Pauli-$Z$ rotations correspond to changing the rotating frame of the qubits, and hence can be implemented virtually.

Comparing Eq.~\eqref{eq:U_B_even_odd} with Eqs.~\eqref{eq:B_odd} and~\eqref{eq:B_even}, the implementation of a B gate using our echoed SDF scheme requires
\begin{equation}
    \delta = \frac{4}{\sqrt{3}} \Omega\,\eta 
    \quad \text{and} \quad
    \Phi = \frac{1}{2}\arccos\!\left[\cot\left(\frac{3\pi}{8}\right)\right]\,.
\end{equation}
which is exactly the result stated in Eq.~\eqref{eq:B_SDF}. The corresponding virtual Pauli-$Z$ rotation angle is
\begin{equation}
    \lambda = \frac{1}{2}\arg\!\left(\cos^2\frac{3\pi}{16} - \sin^2\frac{3\pi}{16}\,e^{-i4\Phi} \right).
\end{equation}

\begin{figure}[t]
    \centering
    \includegraphics[width=0.97 \linewidth]{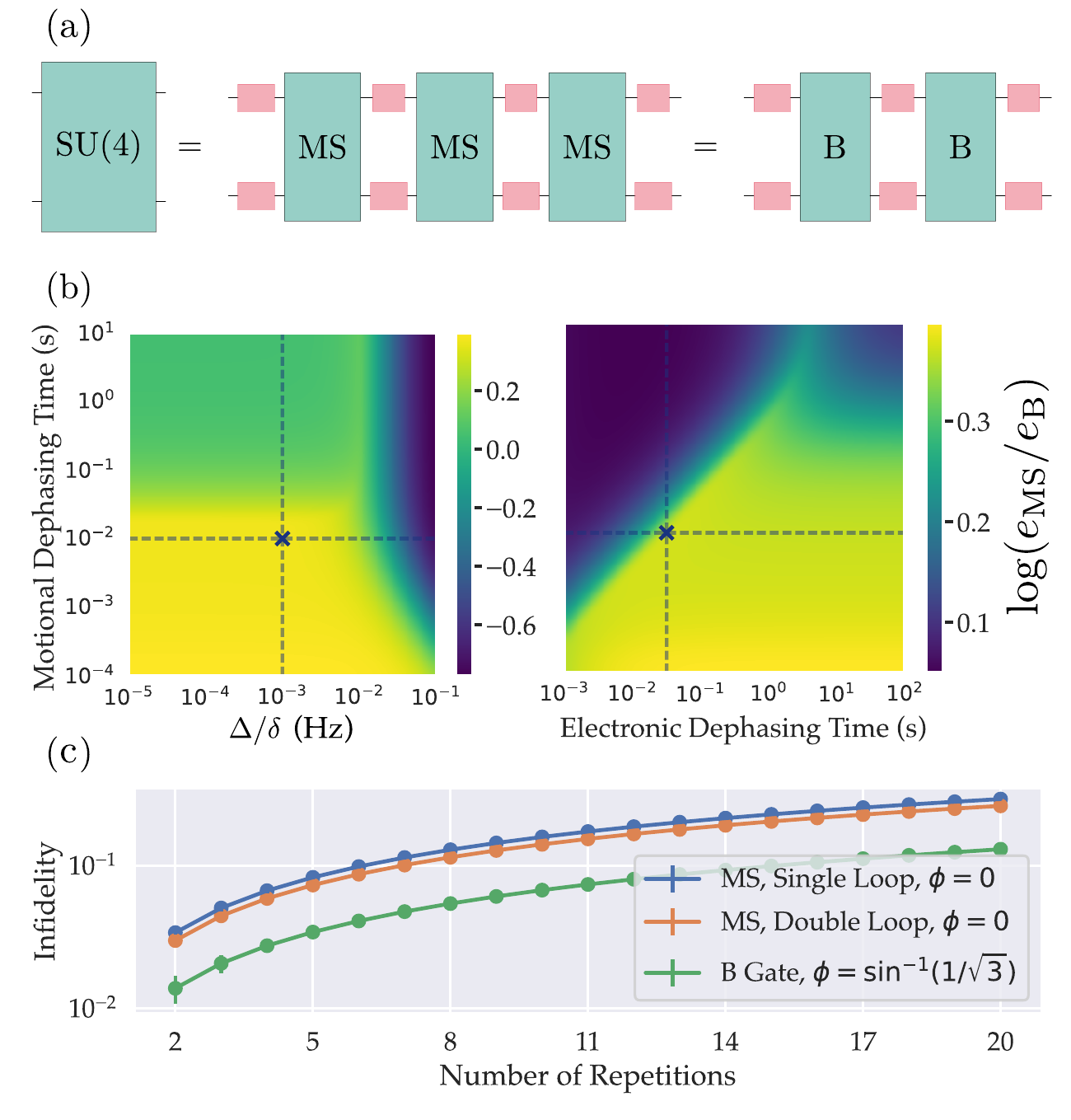}
    \caption{(a) Decomposition of a generic SU(4) two-qubit unitary via MS gates and B gates. (b) Average infidelity comparison for decomposing a randomized arbitrary unitary SU(4) operation with MS-gates ($e_{\text{MS}}$) vs.\ B-gates ($e_{\text{B}}$). As $e_{\text{MS}}$, we take the better performing MS gate implementation between single and double loop \cite{Hayes_2012} variants. Noise parameters: $T_1$=1s, heating rate $\dot n = 1$ quanta/s for all plots; electronic dephasing time (bare $T_2$) set to 50 ms for the left plot and fractional detuning error $\Delta/\delta$ set to $10^{-3}$ for the right plot. Gate times are $83.33 \mu s$, $117.85 \mu s$, $144.3 \mu s$ for the single loop MS-gate, double-loop MS gate and the B-gate, respectively. (c) Infidelity of evolving a two qubit state, averaged over 20 randomizations, through consecutive random SU(4) repetitions. Annotated points from the plots in (b), show the chosen electronic/motional dephasing time and fractional detuning error employed in the bottom plot. Single qubit gates are inserted in between each two qubit gate without noise.}
    \label{fig:fig5}
\end{figure}

\paragraph{B gate simulations --} Decomposition of an arbitrary SU(4) gate in terms of fixed two-qubit gates involves either three MS gates or two B gates (together with arbitrary single-qubit rotations, see Fig.~\ref{fig:fig5}.a) \cite{Zhang_2004}. To compare the performance of both decompositions for trapped ion hardware, we run numerical simulations using \texttt{QuTIP}~\cite{Johansson2012}. To account for the effects of hardware noise, we simulate two (three) successive executions of the B gate (MS gate) using spin-dependent force Hamiltonian in the Lamb-Dicke approximation~\cite{PhysRev.85.259}
\begin{equation} \label{eq:H_SDF}
    H_{\text{SDF}} = \frac{\hbar \Omega \eta}{2} \sigma_+ (a\,e^{i (\delta t - \varphi_r)} + a^{\dagger}\,e^{-i (\delta t + \varphi_b)}) + \text{h.c.},
\end{equation}
where $\Omega$ is the Rabi rate, $\eta$ is the Lamb-Dicke parameter, $a$ ($a^{\dagger}$) denotes the phonon annihilation (creation) operator, $\sigma_+$ the spin raising operator, $\delta$ the laser detuning from sidebands and $\varphi_r$ ($\varphi_b$) the phase of the red (blue) detuned laser (relative to the carrier phase) \cite{RevModPhys.75.281}.

We evolve the system of two spins and motional degrees of freedom using the Lindblad master equation with jump operators
\[
\begin{split}
    L_1 &= \frac{1}{\sqrt{T_1}} (\sigma_1^{-} + \sigma_2^{-}), 
    \quad\,
    L_2 = \sqrt{\frac{1}{2 T_2} - \frac{1}{4 T_1}} (\sigma^Z_1 + \sigma^Z_2), \\
    L_3 &= \sqrt{\dot n} \hspace{0.1cm} a^{\dagger},
    \qquad\qquad\quad\,
    L_4 = \frac{1}{\sqrt{T_m}} a^{\dagger} a
\end{split}
\]
accounting for electronic decay, electronic dephasing, heating and motional dephasing, respectively. Here, $T_1$ and $T_2$ are the characteristic electronic decay and dephasing times, $\dot n$ is the motional heating rate, and $T_m$ is the motional dephasing time. We discretize the total time evolution of $t_g = \frac{2 \pi}{\delta}$ using 50 steps.

We truncate the motional Hilbert space to 20 quanta and use the same Rabi rate $\Omega=100$ kHz and Lamb-Dicke parameter $\eta=0.06$ for all gate variants. We introduce a detuning error $\Delta = |\delta - \delta_{\text{target}}|$ that models imperfect loop closure. We set $\delta_{\text{target}} = 2 \Omega \eta$ for single loop MS-gate, $\delta_{\text{target}} = \sqrt{8} \Omega \eta$ for the double loop MS-gate and $\delta_{\text{target}} = \frac{4 \Omega \eta}{\sqrt{3}}$ for the B-gate. The infidelity of the protocol is quantified by averaging the following quantity over 20 randomized initial states
\begin{equation}
    e = 1- \text{Tr} \left[ \sqrt{\sqrt{\rho_1} \rho_2 \sqrt{\rho_1}} \right]
\end{equation}
where $\rho_1$ ($\rho_2$) is the density matrix of the time evolved state with (without) noise. 

In Fig. \ref{fig:fig5}.b, we compare the performance of MS vs. B gates in decomposing an arbitrary two qubit gate for various values of $T_1, T_2$ and $T_m$. For the simulation of a decomposition using MS gates, we implement a single MS gate both via the single and double loop variants, and plot the better performing variant as its infidelity $e_{\text{MS}}$. In Fig. \ref{fig:fig5}.c, we compare the performance of single loop MS, double loop MS and B gate decompositions of sequential repetitions of randomized two qubit gates. Our implementation is available in the source repository~\cite{erenaykrcn_ccU_2026}.

\iftoggle{arXiv}{
\cleardoublepage 
\title{Supplemental Material:\\Near-Optimal Quantum Time Evolution Circuits \\via Provably Convergent Compression}
\maketitle

\onecolumngrid

\renewcommand{\thefigure}{S\arabic{figure}}
\setcounter{figure}{0}
\renewcommand{\thetable}{S\arabic{table}}
\setcounter{table}{0}
\renewcommand{\theHfigure}{S\arabic{figure}}
\renewcommand{\theHtable}{S\arabic{figure}}
\renewcommand{\theequation}{S\arabic{equation}}
\setcounter{equation}{0}

\vspace{-15pt}

In the main text, we argued how one can guarantee convergence of a variational optimization protocol to a unitary $G_L$ that approximates the time evolution of any local, translationally invariant (TI) Hamiltonian $H$, with a near-optimal scaling given in Eq.~\eqref{eq:approx_err}. Our argument is based on the idea of running this optimization for a short enough target time $\Delta t < t_{\text{crit}}$ and circuit depth $\Delta L$, where this circuit depth $\Delta L$ equals the number of inequivalent permutation classes of connectivity. This number is defined by the system geometry $\mathcal{G}$, that is, the list of all nearest neighbor bonds. Then, as stated in Corollary~\ref{corr:1.1}, repeating this optimized circuit, run for $\Delta t$, until one encodes time evolution for any $t>0$, yields the same favorable scaling from Eq.~\eqref{eq:approx_err}.

\SMsec{Details on the Proof of Theorem 1}
\paragraph{Definition of $H_{\text{opt}}$ --} Here, we show that for a sufficiently small $\Delta t$ the principal matrix logarithm of $G_{\Delta L}$ is well defined, so that $H_{\text{opt}} = \frac{i}{\Delta t} \log G_{\Delta L}$ can be taken from the principal branch. This holds if and only if $-1$ is not in the spectrum of $G_{\Delta L}$, i.e. $-1\notin \sigma(G_{\Delta L})$.

The optimal circuit $G_{\Delta L}$  approximates the target evolution with
\begin{equation} \label{eq:deltat_err}
    \norm{G_{\Delta L} - e^{-i H \Delta t}} \leq \epsilon_0 = \mathcal{O} \left(N \, \Delta t \hspace{0.1cm} e^{- \Omega \left( (\Delta L/\Delta t)^{1/k} \right)}\right)
\end{equation}
for some $k \in \mathbb{N}^+$. Let a sufficiently small $\Delta t$ which simultaneously ensures $\norm{H} \Delta t < \pi - \phi$ for some ${\phi > 0}$ and $\epsilon_0 < 2 \sin(\phi/2)$.

The first condition implies
\begin{equation} \label{eq:spectr_exp_iHDt}
    \sigma(e^{-iH\Delta t}) \subset \{e^{i \lambda} : \lambda \in [-\pi + \phi, \pi - \phi] \}
\end{equation}
i.e., the spectrum of the target is bounded away from -1.
Since both $G_{\Delta L}$ and $e^{-iH\Delta t}$ are unitary, their spectra are stable under operator-norm perturbations~\cite{Bhatia01021984} and the Hausdorff distance between their spectra satisfies
\begin{equation}
    d_H\!\left(\sigma(G_{\Delta L}),\sigma(e^{-iH\Delta t})\right)
\le \norm{G_{\Delta L} - e^{-iH\Delta t}}
\le \epsilon_0.
\end{equation}
Hence, for every eigenvalue $e^{i\psi} \in \sigma(G_{\Delta L})$ there exists $e^{i\theta} \in \sigma(e^{-iH\Delta t})$ with $\theta\in (-\pi + \phi, \pi - \phi]$ (see~Eq.~\eqref{eq:spectr_exp_iHDt}) such that $\abs*{e^{i\psi}-e^{i\theta}} \le \epsilon_0$.

Choosing an arbitrary $\psi\in(\theta-\pi,\theta+\pi]$ and using the identity $\abs{e^{i\psi} - e^{i\theta}} = 2\,\abs{\sin((\psi-\theta)/2)}$ implies that $\abs{\psi-\theta} \le 2\arcsin(\epsilon_0/2)$. By the triangle inequality
\begin{equation}
    \abs{\psi} \le \abs{\theta} + \abs{\psi-\theta} 
    \le (\pi-\phi) + 2\arcsin(\epsilon_0/2)
    = \pi - \tilde\phi,
\end{equation}
where $\tilde\phi := \phi - 2\arcsin(\epsilon_0/2)$. Since $\epsilon_0 < 2\sin(\phi/2)$ ensures $\tilde\phi>0$, we conclude that
\begin{equation} \label{eq:spectr_GDL}
    \sigma(G_{\Delta L})\subset\big\{e^{i\psi}:\psi\in[-\pi+\tilde\phi,\ \pi-\tilde\phi]\big\}\,,
\end{equation}
so $-1\notin\sigma(G_{\Delta L})$ and the principal logarithm $\log(G_{\Delta L})$ is well defined.

\paragraph{Ensuring $\norm{H_0} \leq \norm{H}$ --} Our local Ansatz $W_{\Delta L}$ is a product of ${N \Delta L/2}$ two-qubit gates (each layer containing $N/2$ gates). Given the Magnus expansion bound~\cite{Blanes_2009},
\begin{equation}
    \norm{Z} \leq \sum_{j=1}^{A} \norm{X_j} \quad\text{with}\quad e^{-iZ} = \prod_{j=1}^{A} e^{-i X_j},
\end{equation}
valid whenever $\sum_j \norm*{X_j} < \pi$, we choose initial gate parameters such that $\norm*{H_0^{(j)}} \leq \frac{2}{N \Delta L} \norm*{H}\,\,\forall j$. This ensures
\begin{equation}
    \norm{Z}\equiv \norm{H_0} \Delta t \leq \sum_{j=1}^{N \Delta L/2} \norm{H_0^{(j)}} \Delta t
    \leq \norm*{H} \Delta t.
\end{equation}

\paragraph{Derivation of $t_\mathrm{crit}$ --}
Using the local Lipchitz continuity for the principal matrix logarithm at $G_{\Delta L}$ and $e^{-iH\Delta t}$, we conclude that there exists a constant ${C_{\tilde\phi} < \infty}$ (finite for $\tilde\phi>0$ from Eq.~\eqref{eq:spectr_GDL}) such that
\begin{equation}
\begin{split}
    \norm{H_{\mathrm{opt}}-H} 
    &= \norm{\frac{i}{\Delta t} (\log G_{\Delta L} - \log e^{-iH\Delta t})} \\[5pt]
    &\leq \frac{C_{\tilde\phi}}{\Delta t}\,\norm{G_{\Delta L}-e^{-iH\Delta t}} \leq \frac{C_{\tilde\phi}}{\Delta t}\,\epsilon_0
\end{split}
\end{equation}
with $\epsilon_0$ as defined in Eq.~\eqref{eq:deltat_err}. By the triangle inequality and the condition $\norm*{H_0} \leq \norm*{H}$, we get
\begin{equation}
\begin{split}
    \norm{ H_0 - H_{\text{opt}} } &\leq \norm{H_0 - H} + \norm{H - H_{\text{opt}}}\\
    &\leq 2 \norm{H} + \mathcal{O}(N \, e^{- \Omega ((\Delta L/\Delta t)^{1/k})}).
\end{split}
\end{equation}
Using this bound within Eq.~\eqref{eq:lipschitz_cont_exph0_GL}, we identify $R/t_{\text{crit}} \approx 2 \norm{H}$ up to the correction $\mathcal{O}(N \, e^{- \Omega ((\Delta L/\Delta t)^{1/k})})$. By propagating the correction term, we conclude that the critical time $t_{\text{crit}}$ must follow Eq.~\eqref{eq:t_crit}.

\SMsec{Proof of Lemma 2}

In this section, we prove Lemma~\ref{lemma:lemma_12}, which is pivotal to the proof of Theorem~\ref{theorem:theorem_1}. The proof proceeds in three steps: we first establish that the image of the Ansatz map  $W_{\Delta L}(V)$ is a smooth embedded submanifold $\mathcal{M}\subset\mathrm{SU}(2^N)$ near the identity configuration. We then argue that the optimal unitary $G_{\Delta L}$ lies within $\mathcal{M}$ for sufficiently small $\Delta t$. Finally, we show that the cost function is strictly convex on $\mathcal{M}$ in a neighborhood of $G_{\Delta L}$, guaranteeing convergence of the optimization.

\bigskip
The Ansatz map $W_{\Delta L}(V)$
\begin{equation}
    W_{\Delta L}: \text{SU}(4)^{\Delta L}
    = \underbrace{\text{SU}(4)\times\ldots\times\text{SU}(4)}_{\Delta L} 
    \,\rightarrow\, \text{SU}\!\left(2^N \right)
\end{equation}
has a parameter domain $\text{SU}(4)^{\Delta L}$ which is a smooth manifold of dimension $15\,\Delta L$, equipped with the bi-invariant Riemannian metric inherited from SU(4). We will denote by $\mathfrak{su}(4)$ the Lie algebra of SU(4), i.e., the space of $4\times4$ traceless skew-Hermitian matrices, which has dimension 15. $W_{\Delta L}$ is defined by a circuit construction and is real-analytic, since it is a composition of real-analytic operations (exponential, inclusion, and product maps). This is a generic assumption for most parametrized quantum circuits~\cite{Barzen_2025}.

We design the circuit of the Ansatz $W_{\Delta L}$ so that each of the $\Delta L$ layers acts on a distinct permutation class of nearest-neighbor bonds, meaning no two layers share the same connectivity pattern. Under this condition we now show that the differential $dW_{\Delta L}$ achieves its maximum possible rank $15\,\Delta L$  at the identity configuration  $V_{\mathds{1}}  = (\mathds{1}, \dots, \mathds{1})$.

To see why, consider what happens when we perturb each gate in $V_{\mathds{1}}$ slightly away from the identity, $V_\ell=e^{-iH_\ell\varepsilon}$ for small $\varepsilon$ and  $H_\ell\in\mathfrak{su}(4)$,
\begin{equation}
    W_{\Delta L}\left( e^{-iH_1\varepsilon}, \dots, e^{-iH_{\Delta L} \varepsilon} \right) \approx \mathds{1} - i \varepsilon \,dW_{\Delta L}
    \qquad\text{with}\qquad dW_{\Delta L}=\sum_{\ell=1}^{\Delta L} T_\ell(H_\ell),
\end{equation}
being the differential defined through linear maps $T_\ell : \mathfrak{su}(4) \to \mathfrak{su}(2^N)$ embedding the local generator $H_\ell$ of layer $\ell$ into the full $N$-qubit algebra. The key point is that each layer $\ell$ acts on a different set of bonds (a different permutation class), so the images $T_\ell(\mathfrak{su}(4))$ are linearly independent subspaces of $\mathfrak{su}(2^N)$. Therefore, no perturbation of one layer can be ``compensated" by another, and the differential $dW_{\Delta L}$ has maximal rank \cite{Wiersema_2024}.

Since $dW_{\Delta L}$ is a real-analytic (thus, smooth) function of the gate parameters, and since maximal rank is a stable property under small perturbations, the same full-rank condition holds throughout a ball $\overline{B_\rho(V_{\mathds{1}})}$ of some finite radius $\rho>0$. By the Constant Rank Theorem (see \cite[Thm.~4.12]{LeeISM}), this means the image $\mathcal{M} \coloneqq W_{\Delta L}(\overline{B_\rho(V_{\mathds{1}})})$ is a smooth, embedded submanifold of SU\big($2^N$\big)  of dimension $15\Delta L$. This means that $\mathcal{M}$ is a smooth surface sitting inside the much larger unitary group, with no self-intersections or cusps. Since $\overline{B_\rho(V_{\mathds{1}}))} \subset \text{SU}(4)^{\Delta L}$ is compact and $W_{\Delta L}$ is continuous, its image $\mathcal{M}$ is also compact.

For a sufficiently small $\Delta t$, the optimal $G_{\Delta L}$ lies within $\mathcal{M}$. To show this, we rely on Eq.~\eqref{eq:approx_err} from the main text and proved in Ref.~\cite{Haah_2021},
\begin{equation}
    \norm{G_{\Delta L}-\mathds{1}} 
    \leq \norm{G_{\Delta L}-U_{\Delta t}} +\norm{U_{\Delta t}-\mathds{1}} 
    \leq \epsilon_0 + \norm{H} \Delta t \,.
\end{equation}
For a bounded Hamiltonian $H$, both terms of the upper bound vanish as $\Delta t\rightarrow0$, placing $G_{\Delta L}$ inside $\mathcal{M}$ for a sufficiently small $\Delta t<T_1$. Moreover, we can argue that the optimal unitary $G_{\Delta L}$ is unique within $\mathcal{M}$ for small enough $\Delta t$, as embedded submanifolds have positive reach $\tau > 0$ locally. This implies that $U_{\Delta t} = e^{-iH\Delta t }$ will lie within the tubular neighborhood of $\mathcal{M}$ for $\text{dist}(U_{\Delta t}, \mathcal{M})<\tau$ for all $\Delta t < T_2$, for a $T_2 > 0$. If $U_{\Delta t}$ lies within the tubular neighborhood, then its projection onto the embedded submanifold $\mathcal{M}$ is unique.

We now show that $G_{\Delta L}$ is a strictly convex minimum of the cost function
\begin{equation}
    f(G) \coloneqq \frac{1}{2} \norm{G - U_{\Delta t}}^2
\end{equation}
when $G$ is restricted to move along $\mathcal{M}$. This cost function is equivalent to $J_{\Delta t}(W_{\Delta L}(V))$, given in Eq.~\eqref{eq:cost_func}, up to an additive constant. The main idea is the following: $f(G)$ is the squared (Frobenius) distance from the fixed target $U_{\Delta t}$, which is a perfectly "bowl-shaped" function in the unitary group, as its unrestricted Hessian is simply the identity, $\nabla^2 f = \mathds{1}$. The only reason the cost could develop unwanted curvature (flat directions or local minima) when restricted to $\mathcal{M}$, is if $\mathcal{M}$ itself is curved in a way that distorts the bowl. This curvature is captured by the second fundamental form $\text{II}_G$, a geometric object measuring how much a surface bends inside the ambient space. Concretely, the Riemannian Hessian restricted to $\mathcal{M}$ at any $G\in\mathcal{M}$ satisfies~\cite[Chapter~5]{AbsilMahonySepulchre2008}
\begin{equation}
    \langle \mathrm{Hess}_\mathcal{M}, f(G)[\xi], \xi \rangle 
    = \langle \nabla^2 f(G)[\xi], \xi \rangle + \langle \nabla f(G), \mathrm{II}_G(\xi,\xi)\rangle
    = \norm{\xi}^2 + \langle \nabla f(G), \mathrm{II}_G(\xi,\xi)\rangle
\end{equation}
for any tangent direction $\xi \in T_G \mathcal{M}$. The first term is always positive (from the flat ambient Hessian), while the second term can in principle be negative if the surface curves toward the target point $U_{\Delta t}$. 

At the optimum  $G=G_{\Delta L}$, the gradient $\nabla f(G_{\Delta L}) = G_{\Delta L} - U_{\Delta t}$ points entirely in the normal direction to $\mathcal{M}$ (because $G_{\Delta L}$ is the closest point on $\mathcal{M}$  to $U_{\Delta t}$). This means the potentially harmful second term is controlled by $\norm{G_{\Delta L} - U_{\Delta t}}$. Since the second fundamental form is a bounded tensor on the compact manifold $\mathcal{M}$, there exists a constant $\kappa>0$ such that~\cite{doCarmo1992}
\begin{equation}
    \langle \text{Hess}_{\mathcal{M}} \, f(G_{\Delta L}) [\xi], \xi \rangle  \geq \norm{\xi}^2 (1 - \kappa \norm{G_{\Delta L} - U_{\Delta t}})
\end{equation}
That is, the cost is strictly convex at the optimum as long as the approximation error $\norm{G_{\Delta L}-U_{\Delta t}}$ is not so large that the curvature of $\mathcal{M}$ overwhelms the ``downhill pull" of the cost function. Since this error decays super-polynomially with $\Delta t$, one can always choose $\Delta t<T_3$ small enough to ensure $\kappa\norm{G_{\Delta L}-U_{\Delta t}}<1-\alpha$ for any fixed  $0<\alpha\leq1$, giving
\begin{equation}
    \mathrm{Hess}_\mathcal{M} f(G_{\Delta L}) \geq \alpha\mathds{1}.
\end{equation}

This strict convexity at $G_{\Delta L}$ extends to a finite neighborhood. The Hessian varies smoothly as one moves along $\mathcal{M}$, namely it is Lipschitz continuous with constant $\mathcal{L}>0$
\begin{equation}
    \norm{\text{Hess}_{\mathcal{M}} f(G_1) - \text{Hess}_{\mathcal{M}} f(G_2) } 
    \leq \mathcal{L} \norm{ G_1 - G_2 }\,.
\end{equation}
The positivity of the Hessian is therefore inherited by all points within a ball of radius
\begin{equation}
    R \coloneqq \frac{\alpha}{2\mathcal{L}}
\end{equation}
around $G_{\Delta L}$ on $\mathcal{M}$. Within this ball $B_{R}(G_{\Delta L}) \, \cap \mathcal{M}$, the cost function has no spurious local minima and any gradient-based optimizer is guaranteed to converge to $G_{\Delta L}$.

Crucially, neither $\alpha$ nor $\mathcal{L}$ depend on $\Delta t$, once we assume that $\Delta t < T = \min (T_1, T_2, T_3)$ such that the following assumptions hold altogether: $\mathcal{M}$ is an embedded submanifold ($\Delta t < T_1$), $U_{\Delta t}$ lies within the tubular neighborhood of $\mathcal{M}$ ($\Delta t < T_2$), and curvature of $\mathcal{M}$ is not steep enough to change the sign of the Hessian eigenvalues ($\Delta t < T_3$). Then, these constants $\alpha$ and $\mathcal{L}$, hence also $R$, can be defined independent of $\Delta t$.

\SMsec{Details on Trotter Circuits for Controlled Evolution}

Both for the numerical benchmarks reported in Figs.~\ref{fig:fig3} and \ref{fig:fig6} and for the initialization of the TICC optimization, we use a $p^{\text{th}}$-order Trotterization of the target Heisenberg Hamiltonian in a field \eqref{eq:HM}, on various different geometries, with the anti-commuting Pauli strings insertion method, proposed and discussed extensively in prior works \cite{ticc, qetu, Babbush_2018}.

This method exploits the well-known equivalence between globally controlled time evolution of time $t$ via a single ancilla qubit and a circuit sequence where the ancilla acts as a control switch between $\pm t/2$ evolutions
\begin{equation}
\label{eq:equivalence}
    \ketbra{0}_{\text{anc}} \otimes \mathds{1} + \ketbra{1}_{\text{anc}} \otimes e^{-i H t} \iff 
    \ketbra{0}_{\text{anc}} \otimes e^{i H t/2} + \ketbra{1}_{\text{anc}} \otimes e^{-i H t/2}.
\end{equation}
Such a circuit equivalence holds within each energy eigenspace of $H$, differing only by a global phase factor. In algorithms where one is interested in only a single energy subspace, e.g. ground-state preparation or ground-state energy estimation, one can replace the globally controlled evolution with a circuit in which the ancilla controls the evolution direction. Notable examples of algorithms where the controlled evolution can be implemented in this way include Quantum Phase Estimation (not only the iterative variant \cite{rpe} but also the full Quantum Fourier Transform appended variant, as shown in Ref.~\cite{ticc}) and eigenspace filtering methods \cite{qetu, aqcf}.

This equivalence reduces overhead in implementing the controlled evolution sequence by avoiding the control of each two-qubit gate in the Trotterized circuit. Instead, one can partition the Hamiltonian $H$ into a set of sub-Hamiltonians $H = \sum_iH_i$, for each of which one can find easy-to-control unitaries $K_i$ (e.g. Pauli strings), such that 
\begin{equation}
    K_i^{\dagger} H_i K_i = - H_i
\end{equation}
holds for each $H_i$. Then, one can choose an appropriate Trotterization of $H$ into these sub-Hamiltonians $H_i$, and insert the anti-commuting $K_i$ unitaries before and after each block $e^{-i H_i t}$. Controlling only these layers $K_i$ with an ancilla implements the circuit equivalence given in Eq.~\eqref{eq:equivalence}. 

For our target Hamiltonian, defined in Eq.~\eqref{eq:HM}, we choose the following partition of the Pauli terms in the Hamiltonian with the corresponding Pauli strings that anti-commute with each of them
\begin{equation}
    H_1 = \sum_{\langle i, j \rangle} \sigma^X_i \sigma^X_j  \hspace{0.1cm},  K_1 = \bigotimes_{i=1}^{N/2}  (\sigma^Z \otimes \mathds{1}), \hspace{1cm}
    H_2 = \sum_{\langle i, j \rangle} \sigma^Y_i \sigma^Y_j  \hspace{0.1cm},  K_2 = \bigotimes_{i=1}^{N/2}  \left(\sigma^X \otimes \mathds{1}\right),
\end{equation}\\[-20pt]
\begin{equation}
    H_3 = \sum_{\langle i, j \rangle} \sigma^Z_i \sigma^Z_j  \hspace{0.1cm},  K_3 = \bigotimes_{i=1}^{N/2}  (\sigma^Y \otimes \mathds{1}), \hspace{1cm}
    H_4 =  - \sum_{i=0}^{N-1} \sigma^Y_i \hspace{0.1cm},  K_4 = \bigotimes_{i=1}^{N/2}  \left(\sigma^X \otimes \sigma^X\right), 
\end{equation}\\[-20pt]
\begin{equation}
    H_5 = \sum_{i=0}^{N-1} \sigma^Z_i \hspace{0.1cm}, K_5 = \bigotimes_{i=1}^{N/2}  \left(\sigma^X \otimes \sigma^X\right), \hspace{1cm}
    H_6 = 3 \sum_{i=0}^{N-1} \sigma^X_i \hspace{0.1cm}, K_6 = \bigotimes_{i=1}^{N/2} \left(\sigma^Z \otimes \sigma^Z\right).
\end{equation}
For one-dimensional circuits, it is enough to implement each $\{e^{-i H_i t}\}_{i=1}^6$ via two-qubit gates and insert the corresponding Pauli strings $K_i$ before and after. For higher-dimensional geometries, each two-body sub-Hamiltonian $H_i$ must be further decomposed into $d$ inequivalent permutation classes associated with the lattice symmetries $H_i = \sum_{\alpha=1}^d H_i^{(\alpha)}$, such that $\big(K^{(\alpha)}_i\big)^{\dagger} H^{(\alpha)}_i K^{(\alpha)}_i = - H^{(\alpha)}_i$. Although all of the $H^{(\alpha)}_i$ terms originate from the same Pauli sector, a difference in connectivity requires a difference in connectivity in the $K^{(\alpha)}_i$ terms. Hence, one has to insert such $K^{(\alpha)}_i$ layers, before and after each $e^{-iH_i^{(\alpha)} t}$ term. This additional decomposition with respect to permutation classes only applies to the two-body terms ($H_1$, $H_2$, $H_3$) as single-body terms are independent of neighbor connectivity. 
For higher order Trotterization $p>1$, each of the blocks $H_i^{(\alpha)}$ is repeated with the corresponding Trotter coefficient and according to the corresponding ordering.

\begin{figure}[b]
    \centering
    \includegraphics[width=0.80\linewidth]{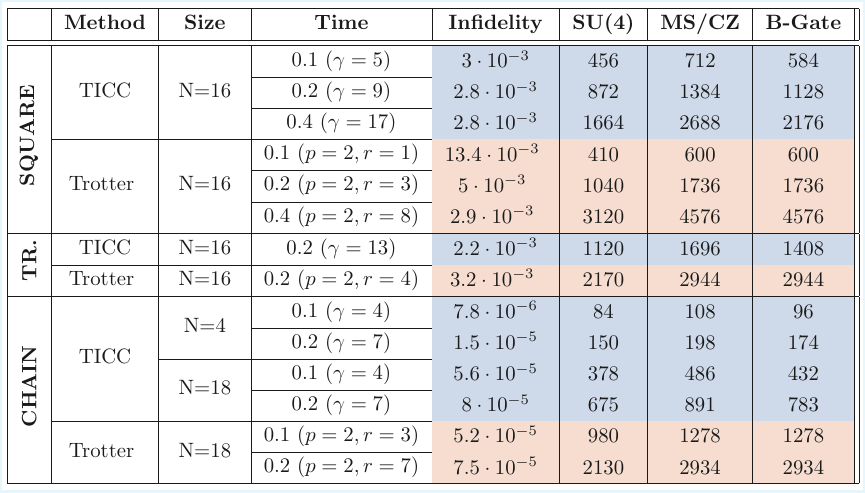}
    \caption{Evolution infidelity \eqref{eq:ev_infid} vs.\ gate counts needed to encode the controlled time evolution of the HM \eqref{eq:HM} on various geometries: 4$\times$4 square lattice, 4$\times$4 triangular lattice, and one-dimensional chain. TICC with control layers $\gamma$ is benchmarked against  $p^{\text{th}}$-order Trotterization (of $r$ steps) with anti-commuting Pauli strings insertion. Optimization with TICC yields arbitrary SU(4) gates, which we decompose into MS and B gates. One-dimensional results are optimized on $N=4$ qubits and transferred to $N=18$ qubits. We report evolution infidelities for both $N=4$ and $N=18$ systems for TICC to demonstrate the linear boundedness of the error with respect to system size. }
    \label{fig:fig6}
\end{figure}

Moreover, the meaning of ``inequivalent permutation classes'' has been discussed in Ref.~\cite{ticc}. We use the same definition as there and explicitly give the inequivalent classes of nearest-neighbor permutations for the considered geometries. We use
\begin{equation}
\begin{split}
    \text{perms}_1 = \big[&(0, 4, 1, 5, 2, 6, 3, 7, 8, 12, 9, 13, 10, 14, 11, 15), \\[3pt] 
                          &(4, 8, 5, 9, 6, 10, 7, 11, 12, 0, 13, 1, 14, 2, 15, 3)\big],
\end{split}
\end{equation}\\[-20pt]
\begin{equation}
\begin{split}
    \text{perms}_2 = \big[&(0, 1, 4, 5, 8, 9, 12, 13, 2, 3, 6, 7, 10, 11, 14, 15), \\[3pt]
                          &(1, 2, 5, 6, 9, 10, 13, 14, 3, 0, 7, 4, 11, 8, 15, 12)\big]
\end{split}
\end{equation}
for the 4$\times$4 square lattice (number of inequivalent classes $d=2$),
\begin{equation}
\begin{split}
    \text{perms}_1 = \big[&(0, 1, 2, 3, 4, 5, 6, 7, 8, 9, 10, 11, 12, 13, 14, 15), \\[3pt] 
                          &(1, 2, 3, 0, 5, 6, 7, 4, 9, 10, 11, 8, 13, 14, 15, 12)\big]
\end{split}
\end{equation}\\[-20pt]
\begin{equation}
\begin{split}
    \text{perms}_2 = \big[&(0, 5, 10, 15, 3, 4, 9, 14, 2, 7, 8, 13, 1, 6, 11, 12), \\[3pt] 
                      &(5, 10, 15, 0, 4, 9, 14, 3, 7, 8, 13, 2, 6, 11, 12, 1)\big] 
\end{split}
\end{equation}\\[-20pt]
\begin{equation}
\begin{split}
    \text{perms}_3 = \big[&(0, 4, 8, 12, 1, 5, 9, 13, 2, 6, 10, 14, 3, 7, 11, 15), \\[3pt] 
                          &(4, 8, 12, 0, 5, 9, 13, 1, 6, 10, 14, 2, 7, 11, 15, 3)\big]
\end{split}
\end{equation}
for the 4$\times$4 triangular lattice (number of inequivalent classes $d=3$) and \begin{equation}
    \text{perms}_1 = \big[(0, 4, 6, 10, 2, 5, 8, 11), (4, 6, 10, 0, 5, 8, 11, 2)\big], \hspace{0.7cm}
    \text{perms}_2 = \big[(0, 1, 2, 3, 6, 7, 8, 9), (1, 2, 3, 0, 7, 8, 9, 6)\big] 
\end{equation}\\[-30pt]
\begin{equation}
    \text{perms}_3 = \big[(1, 4, 9, 11, 3, 5, 7, 10), (4, 1, 11, 9, 5, 7, 10, 3)\big]
\end{equation}
for the $N=12$ Kagome lattice (number of inequivalent classes $d=3$). In this notation, each pair of consecutive entries $(a_{2j}, a_{2j+1})$ within a permutation denotes a nearest-neighbor bond. Each such permutation corresponds to one circuit layer, because two-qubit gate operations that connect each neighbor of the given permutation can be parallelized. These three inequivalent permutation classes of the Kagome lattice are visualized in Fig.~\ref{fig:fig1}.

\SMsec{Numerical Benchmarking of TICC on Additional Geometries}

To further validate the performance of TICC and the idea of bootstrapping shallow-depth and short-time-optimized circuits to encode large times, we run optimizations on additional lattice geometries of the same Hamiltonian from Eq.~\eqref{eq:HM}. These results are reported in Fig.~\ref{fig:fig6}.

\SMsec{Details on Gate Synthesis}

For the gate counts given in Fig.~\ref{fig:fig3} and Fig.~\ref{fig:fig6}, we use the approximate gate synthesis library \texttt{BQSKit}~\cite{doecode_58510}. We set an approximation error upper bound of \texttt{synthesis\_epsilon=1e-6} for decomposing the set of controlled two-qubit gates, employed in the control layers of TICC, into CZ/B gates. This decomposition results in $\sim 9$ CZ gates, per controlled two-qubit gate, on average. The gates in the uncontrolled layers of TICC, as well as the gates of Trotter circuits, were decomposed into CZ and B gates via exact decomposition, corresponding to machine precision in synthesis error.

\SMsec{Additional Verification of Transferability using Pauli Propagation}

In order to additionally validate the transferability of the optimized circuits, run on the $N=12$ Kagome lattice, we use Pauli Propagation framework \cite{rudolph2025paulipropagationcomputationalframework} to evaluate evolution infidelities on larger systems, supporting our PEPS simulation results.

According to Ref. \cite{caro2023out}, one can sample states $v$ from the single qubit Haar measure $ v\sim \mathrm{Haar}_1^{\otimes n}$ instead of the global Haar measure, for computing the cost function given in Eq.~\eqref{eq:C_hst}. This modified cost function $C_t^1(W_L(V))$ constitutes an upper bound to $C_t(W_L(V))$ by
\begin{equation}
    \frac12 C_t(W_L(V)) \leq \frac{2^N}{2^N+1} C_t^1(W_L(V)),
\end{equation}
where
\begin{equation}
    C_t^1(W_L(V)) \coloneqq  1 - \mathbb{E} \left[ \abs*{\braket{ v | U_t^{\dagger} W_L(V) | v}}^2 \right]_{v\sim\mathrm{Haar}_1^{\otimes n}}.
\end{equation}
Moreover, authors of Ref. \cite{danna2025circuitcompression2dquantum} showed that $C_t^1(W_L(V))$ can be transformed into a local cost function 
\begin{equation}\label{eq:PPcost}
    C_t^{1,\mathrm{loc}}(W_L(V)) \coloneqq \frac12 - \frac1{6N}\sum_{j=1}^N \sum_{\sigma=\sigma^X, \sigma^Y, \sigma^Z} \frac1{2^N}\Tr\left[U_t^\dagger \sigma_j U_t W_L(V)^\dagger \sigma_j W_L(V)\right],
\end{equation}
which is easier to compute compared to both $C_t(W_L(V))$ and $C_t^1(W_L(V))$. This local cost function $C_t^{1,\mathrm{loc}}(W_L(V))$ gives a ``local infidelity'' $\mathcal I^{\mathrm{loc}}$ that upper bounds $C_t$
\begin{equation}
\label{eq:epsilon_max}
    C_t(W_L(V))  \leq \mathcal I^{\mathrm{loc}} \, \coloneqq \, 2N\frac{2^N}{2^N +1} C_t^{1,\mathrm{loc}}(W_L(V)).
\end{equation}
Hence, upper bounding the cost function $C_t(W_L(V))$ only requires sampling the evolution of weight-1 Pauli strings $\Tr\left(U_t^\dagger \sigma_j U_t W_L(V)^\dagger \sigma_j W_L(V)\right)$, which we perform using the Pauli propagation (PP) framework \cite{rudolph2025paulipropagationcomputationalframework}.

\begin{figure}[t]
    \centering
    \includegraphics[width=0.95\linewidth]{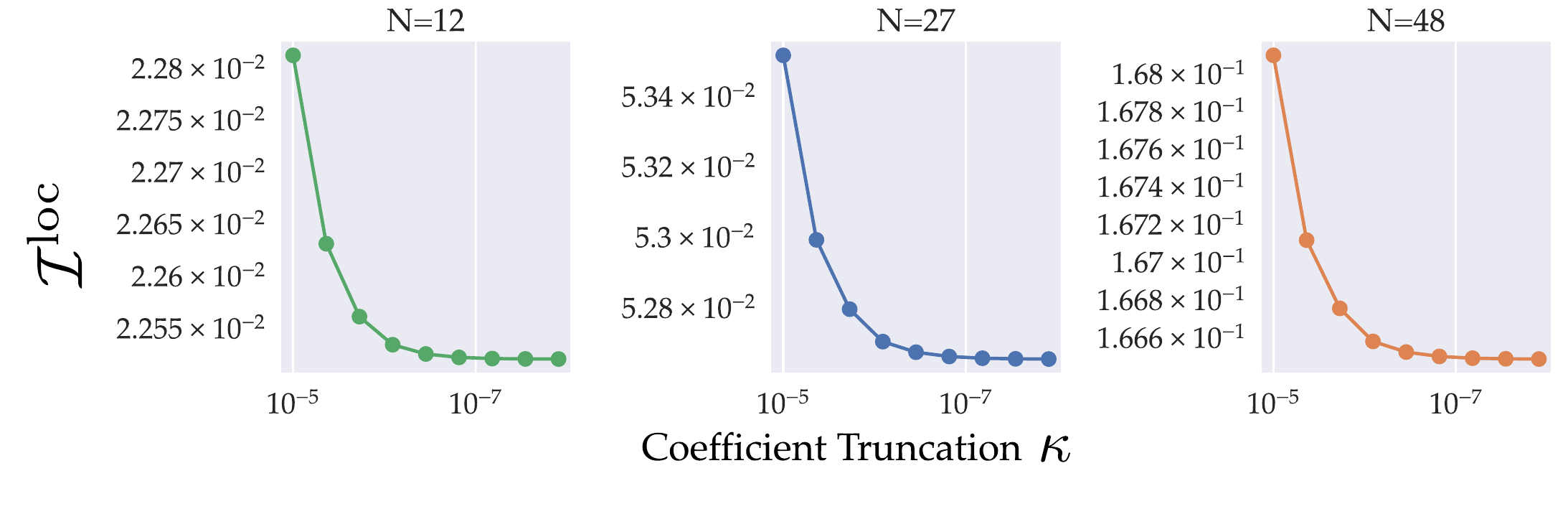}
    \caption{Convergence of local infidelity $\mathcal I^{\mathrm{loc}}$, defined in Eq.~\eqref{eq:epsilon_max}, as a function of coefficient truncation $\kappa$ for the Kagome lattice using TICC for $N=12$ (left). Additionally, we estimate $\mathcal I^{\mathrm{loc}}$ for the larger lattices $N=27$ (center) and $N=48$ (right) by transferring the optimized circuit.}
    \label{fig:PPcost}
\end{figure}

PP is a Heisenberg picture method that expands observables in the Pauli basis $P\in\{\mathds{1}, \sigma^X, \sigma^Y, \sigma^Z\}^{\otimes N}$
\begin{equation}\label{eq:PPexpansion}
    U^\dagger P_0 U = \sum_P \alpha_P P,
\end{equation}
so that the scalar product $\Tr\left(U^\dagger \sigma_j U V^\dagger \sigma_j V\right)$ is given by
\begin{equation}
\label{eq:Pauli_strings}
    \Tr\left(U_t^\dagger \sigma_j U_t W_L(V)^\dagger \sigma_j W_L(V)\right) = \sum_{P, P'} \alpha_P\beta_{P'} \Tr(P P') = \sum_P \alpha_P\beta_{P}
\end{equation}
where $U_t^\dagger \sigma_j U_t \equiv \sum_P \alpha_P P$ and $W_L(V)^\dagger \sigma_j W_L(V) \equiv \sum_{P'} \beta_{P'} P'$. The last step in Eq.~\eqref{eq:Pauli_strings} follows from the orthogonality of the Pauli strings.

The expansion given in Eq.~\eqref{eq:PPexpansion} can include exponentially many terms, so in order to keep the method classically tractable, it is necessary to truncate the sum. The most common truncation is coefficient truncation, where one sets a cut-off $\kappa$ and discards all strings with $|\alpha_P| < \kappa$.

In Fig. \ref{fig:PPcost} we evaluate the ``local infidelity'' $\mathcal I^{\mathrm{loc}}$ given in Eq.~\eqref{eq:epsilon_max} for the evolution of the HM given in Eq.~\eqref{eq:HM}, on Kagome lattices of various system sizes. This time evolution circuit was optimized on the $N=12$ system, using TICC, whose optimized gates were reused to encode larger system sizes $N=27$ and $N=48$. We run PP with various coefficient truncation cut-off values $\kappa$, and demonstrate convergence for all considered systems. The local infidelities we obtain with PP are in good agreement with what we found using PEPS (see bottom panel of Fig. \ref{fig:fig3}).

}

\end{document}